\documentclass[pre,aps]{revtex4}

\usepackage{graphicx}

\begin{document}

\title{Propagation of Bessel beams
 in absorbing media:\\
   a new generation of GPR devices?}
\author{D. Mugnai and P. Spalla}
\affiliation{``Nello Carrara'' Institute of Applied Physics,  CNR
Florence Research Area,\\
 Via Madonna del Piano 10, 50019 Sesto
Fiorentino (FI), Italy}


 \maketitle

\newcommand{\be}{\begin{equation}}
\newcommand{\ee}{\end{equation}}
\newcommand{\bea}{\begin{eqnarray}}
\newcommand{\eea}{\end{eqnarray}}

In recent years localized waves have aroused great interest in the
scientific community \cite{book}. In relation to this topic, many
efforts were devoted to the analysis of Bessel beams because of
their unusual features: they are non-diffracting
\cite{dur87,zio,spr,saa,tan} and show superluminal behavior both
in phase and group velocities \cite{mug00,ale,bes,zam,saa1}.

The aim of this paper is the study of the propagation of a Bessel
beam through two absorbing layers, limited by two different
half-spaces. Our approach will be based on the scalar analysis,
since this analysis was proved to be an excellent approximation of
the vectorial field which describes a Bessel beam \cite{mug06}.

\vspace{0.5 cm}

\begin{figure}
\includegraphics[scale=0.6]{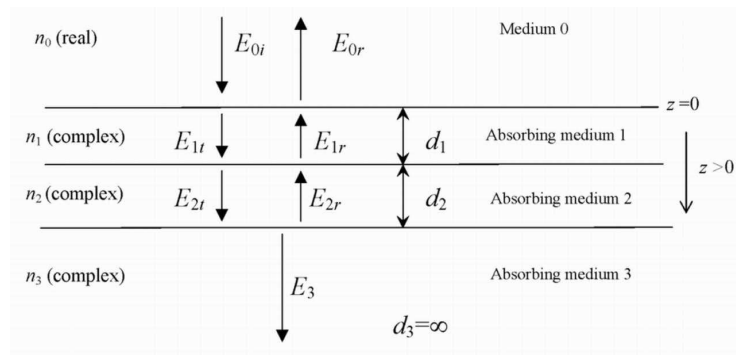}
\caption{Schematic representation of propagation. A Bessel beam
impinges on medium 1 at normal incidence, then in medium 2 and,
lastly, propagates through medium 3. Media 1,2 and 3 are absorbing
media.} \label{scheme}
\end{figure}

Let us consider a system formed by two layers, 1 and 2, limited by
two half-spaces denoted by 0 and 3,  as sketched in Fig.
\ref{scheme}.  

All planes limiting the four different media are parallel and of
infinite length. A localized wave, at frequency $\omega$, impinges
on medium 1 coming from medium 0, which is non absorbing. On the
contrary, media 1, 2, and 3 are absorbing. Be $d_1$ and $d_2$ the
thickness of the layer 1 (medium 1) and 2 (medium 2) respectively.
Let us denote with the subscripts $i,\:r$ and $t$  the incident,
reflected and transmitted fields, respectively. The subscripts 0,
1, 2, and 3 characterized the media in which the field propagates.
Be $n_0,\:n_1, \:n_2,\:n_3$ the refractive indices of media 0, 1,
2, 3, respectively. For the time being, let us consider an ideal
perfect homogeneous system.

Let us consider a localized wave of Bessel type which impinges in
medium 1 at normal incidence. We wish to remind that a Bessel wave
- or Bessel beam - originates by the interference of an infinite
number of plane waves,  each of which tilted of the same angle
$\theta_0$ with respect to a given direction, say $z$, which is
the direction of propagation of the beam.
 We refer to a specific Bessel beam, namely the one described
in Ref. \cite{mug06}. For a Bessel beam like the one described in
this particular reference, the field is linearly polarized and the
scalar approximation is justified since two components of the
field are negligible as compared  to the third component, which is
the dominant one.  With reference to a cylindrical coordinate
system ($\rho , \:\psi , z$), a Bessel beam $E$ is given by

\be
 E =  J_0(nk_0\rho\sin\theta_0 \,)e^{i n k_0  z\cos\theta_0},
 \label{bessel}
 \ee
where $J_0$ is the Bessel function of first type, $\theta_0$ is
the axicone angle which characterizes the Bessel beam, $n$ is the
refractive index of the medium, and $k_0 =\omega /c$ is the wave
number in vacuum (the field $E$ is rotationally symmetric and thus
independent on the angular coordinate $\psi$). Thus, the electric
fields in media 0, 1, 2, 3 can be written as (see Fig.
\ref{scheme})

\bea
 E_{0i}&=& J_0\,e^{ik_0n_0z\cos\theta_0},  \hspace{1 cm}      (z\leq 0)
  \label{e0i}  \\
  E_{0r}&=& r_0\, J_0 \,e^{-
  ik_0n_0z\cos\theta_0} , \hspace{1 cm}   (z\leq 0)
  \label{e0r}  \\
 E_{1t}&=& t_1\, J_{01} \, e^{ ik_0n_1
 z\cos\theta_1},  \hspace{1 cm}     (0\leq z\leq d_1)
  \label{e1t}  \\
   E_{1r}&=& r_1 \,J_{01} \, e^{- ik_0n_1 z\cos\theta_1},  \hspace{1 cm}  (0\leq z\leq d_1)
  \label{e1r}  \\
   E_{2t}&=& t_2 \,J_{02}\, e^{ ik_0n_2 (z-d_1) \cos\theta_2}, \hspace{1 cm}   (d_1\leq z\leq (d_1+d_2)
  \label{e2t}  \\
    E_{2r}&=& r_2\, J_{02}\, e^{- ik_0n_2 (z-d_1) \cos\theta_2},  \hspace{1 cm}  (d_1\leq z\leq (d_1+d_2)
  \label{e2r}  \\
  E_{3}&=& t_{3}\, J_{03}\, e^{ ik_0n_3 (z-(d_1+d_2)) \cos\theta_{3}}, \hspace{1 cm} ( z \geq (d_1+d_2))
  \label{e3}
 \eea
where
\bea
 J_0=J_0(k_0n_0\rho\sin\theta_0 ), \:\:J_{01}=J_0(k_0n_1\rho\sin\theta_1
 ),\:\: J_{02}=J_0(k_0n_2\rho\sin\theta_2 )\:\:
 J_{03}=J_0(k_0n_3\rho\sin\theta_{3}) ,
 \nonumber
 \eea
and  $\theta_1, \: \theta_2$, and $\theta_{3}$ are the complex
angles which characterize the Bessel function inside the media 1,
2 and 3, respectively. The temporal factor $e^{-i\omega t}$, which
is present in all fields, is omitted for the sake of simplicity.

All the refractive indices are complex, with the exception of
$n_0$. The quantities $r,\: t$, which are labeled with reference
to the medium in which the field propagates, denote the complex
reflection and transmission coefficients, respectively.

It is crucial to note that, even if the refractive index $n_1$ and
the angle $\theta_1$ are both complex, the product
$n_1\sin\theta_1$ is real \cite{mugnopub}. In fact, in the
propagation of a plane wave in a conducting medium, the surfaces
of constant amplitude are parallel to the interface \cite{str}.
Thus, in order to meet this condition the product
$n_1\sin\theta_1$  has to be real, otherwise the
constant-amplitude surfaces would be plane, but with an
inclination with respect to the interface. By considering the
continuity of the phase of the incident and reflected waves, it is
also possible to demonstrate that $n_1\sin\theta_1=
n_0\sin\theta_0$ \cite{mugnopub}.

Since all the products $n_n\sin\theta_n$ are real and are equal to
$n_0\sin\theta_0$, we come to the remarkable conclusion that the
shape of the Bessel beam is not modified when propagating into an
absorbing medium. As a consequence, Eqs. (\ref{e0i})-(\ref{e3})
can  be written as

\bea
 E_{0i}&=& J_0\,e^{ik_0n_0z\cos\theta_0},
  \label{e0im}  \\
  E_{0r}&=& r_0\, J_0 \,e^{- ik_0n_0z\cos\theta_0} ,
  \label{e0rm}  \\
  E_{1t}&=& t_1\, J_{0} \, e^{ ik_0n_1  z\cos\theta_1},
  \label{e1tm}  \\
   E_{1r}&=& r_1 \,J_{0} \, e^{- ik_0n_1 z\cos\theta_1},
  \label{e1rm}  \\
   E_{2t}&=& t_2 \,J_{0}\, e^{ ik_0n_2 (z-d_1) \cos\theta_2},
  \label{e2tm}  \\
    E_{2r}&=& r_2\, J_{0}\, e^{- ik_0n_2 (z-d_1) \cos\theta_2},
  \label{e2rm}  \\
  E_{3}&=& t_{3}\, J_{0}\, e^{ ik_0n_3 (z-(d_1+d_2)) \cos\theta_{3}}.
  \label{e3m}
 \eea

In order to evaluate the reflection and transmission coefficients,
we shall apply the boundary conditions to the field (that is the
condition of continuity of the field on the border surface between
two media), and to its first derivative with respect to the
direction of propagation. The latter condition works only in the
scalar approximation, and is none other than the condition of
continuity of the tangential component of the magnetic field.
 In formula we have:

  \bea
&& [E_{0i}+E_{0r}]_{z=0} = [E_{1t}+ E_{1r}]_{z=0} \nonumber \\
 && \left[E_{1t}+E_{1r}\right]_{z=d_1} = [E_{2t}+E_{2r}]_{z=d_1} \label{co-field} \\
 && \left[E_{2t}+E_{2r}\right]_{z=(d_1+d_2)} =  [E_{3}]_{z=(d_1+d_2)} \nonumber \:\: ,
  \eea
and

 \bea
 && \left[\frac{\partial{E_{0i}}}{\partial z}+ \frac{\partial{E_{0r}}}{\partial z}\right]_{z=0}
 = \left[\frac{\partial{E_{1t}}}{\partial z}+ \frac{\partial{E_{1r}}}{\partial z}\right]_{z=0} \nonumber \\
 && \left[\frac{\partial{E_{1t}}}{\partial z}+\frac{\partial{E_{1r}}}{\partial z}\right]_{z=d_1}
 = \left[\frac{\partial{E_{2t}}}{\partial z}+ \frac{\partial{E_{2r}}}{\partial z}\right]_{z=d_1}
  \label{co-der} \\
 && \left[\frac{\partial{E_{2t}}}{\partial z}+ \frac{\partial{E_{2r}}}{\partial z}\right]_{z=(d_1+d_2)}
  =\left[\frac{\partial{E_{3}}}{\partial
  z}\right]_{z=(d_1+d_2)}\nonumber .
   \eea
With the notations

\begin{figure}[t]
\includegraphics[scale=0.5]{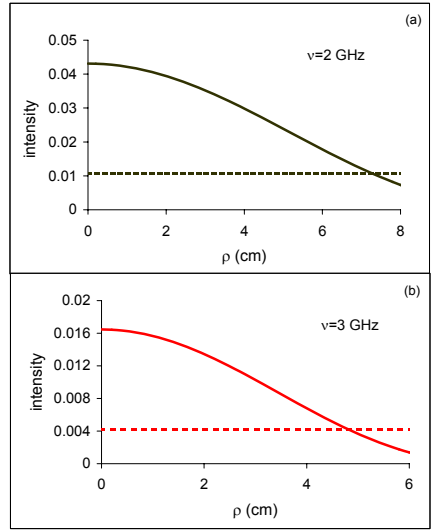}
\caption{Intensity of the reflected field in medium 0 (Eq.
\ref{e0r}), as a function of the radial coordinate $\rho$. Near
$\rho\simeq 0$ the intensity of the Bessel beam is greater than
that of the plane wave. On going away from the origin, the
intensity of the beam decreases and the intensity of plane wave
surmounts. Parameter values are: $\theta_0=30^{\circ} , \:
n_{1r}=1.1,\: n_{2r}=2.2, d_1=25$ cm, $d_2=1$ cm,$ \:\nu =2$ GHz
(a), and $\nu =3$ GHz (b). The values of the refractive indices,
$n_{1r}$ and $n_{2r}$, refer to dry sandy ground (medium 1) and
glass (medium 2), respectively. The $n_{2r}=2.2$ value may also
refer to other kinds of materials, such as stone or ceramics
materials}
 \label{ro}
\end{figure}

 \bea
&& \alpha_1=k_0n_1d_1\cos\theta_1,
\;\:\:\alpha_2=k_0n_2d_2\cos\theta_2, \nonumber \\
  && \phi_0=k_0n_0\cos\theta_0 ,\:\:\:
  \phi_1=k_0n_1\cos\theta_1,
 \:\:\: \phi_2=k_0n_2\cos\theta_2, \:\:\:
 \phi_{3}=k_0n_3\cos\theta_{3},
 \eea
 Eqs. (\ref{co-field}) and (\ref{co-der}) can be written as:
 \bea
&& 1+r_0 = t_1+r_1 \nonumber
\\
 &&  t_1e^{i\alpha_1} + r_1e^{-i\alpha_1}  = t_2+r_2   \label{field} \\
 &&   t_2e^{i\alpha_2}+r_2e^{-i\alpha_2}  =  t_{3}  \nonumber ,
  \eea
   \bea
&& \phi_0(1-r_0) =  \phi_1(t_1-r_1) \nonumber \\
 &&  \phi_1 \left(t_1e^{i\alpha_1} - r_1e^{-i\alpha_1} \right) = \phi_2 (t_2-r_2 ) \label{der} \\
 &&  \phi_2 \left( t_2e^{i\alpha_2}-r_2e^{-i\alpha_2} \right) =  t_{3}\phi_{3}  \nonumber  .
  \eea
 By solving equations (\ref{field}) and (\ref{der}), we
 obtain the reflection coefficient $r_0$ related to the propagation in the incoming medium
 0, and the reflection and transmission coefficients related to the
propagation in the layers 1,2 and in the half-space 3 (the
calculations are rather laborious but present no difficulty)

\bea
 r_0 &=&\frac{a\, \phi_0- a_1\,\phi_1  }{a\, \phi_0+ a_1\,\phi_1  },
 \label{r0} \\
 r_1&=&
 \frac{ 1+r_0-r_2A^-e^{-i\alpha_1} }
 {1-e^{-2i\alpha_1}}  \label{r1}\\
r_2&=&-\:\frac{2\phi_1 e^{-i\alpha_1}(1+r_0)}
 {\phi_2A^+\left(e^{-2i\alpha_1}-1\right)-\phi_1A^-\left(e^{-2i\alpha_1}+1\right)}\label{r2}\\
  t_1&=& e^{-i\alpha_1} \left(r_2A^- -r_1  e^{-i\alpha_1}\right)\label{t1}\\
 t_2&=&-r_2e^{-2i\alpha_2}  \Phi \label{t2}\\
 t_3&=& \frac{\phi_2}{\phi_3}
 \left(t_2e^{i\alpha_2}-r_2e^{-i\alpha_2}\right) \label{t3},
 \eea
 where
\bea
 a &=& (e^{-2i\alpha_1}-1)^2(\phi_1A^-+\phi_2A^+)-2\phi_1A^- e^{-2i\alpha_1}(e^{-2i\alpha_1}-1)
 \nonumber \\
 a_1 &=& \left[\frac{}{}(e^{-2i\alpha_1}-1)(\phi_1A^-+\phi_2A^+)-2\phi_1 A^-e^{-2i\alpha_1}\right] (e^{-2i\alpha_1}+1)+4\phi_1A^-
 e^{-2i\alpha_1}\nonumber \\
 \Phi &=& \frac
 {\phi_{3}+\phi_2}{\phi_{3}-\phi_2} \nonumber
 \eea
and \bea
 A^+ = 1+e^{-2i\alpha_2} \Phi,
  \;\;\;\: A^- = 1-e^{-2i\alpha_2}\Phi . \nonumber
 \eea
By replacing  Eqs. (\ref{r0})-(\ref{t3}) in Eqs.
(\ref{e0im})-(\ref{e3m}), we are now able to obtain the
electromagnetic fields that describe the propagation of a Bessel
beam through the multilayer system of Fig. \ref{scheme}.

\vspace{1 cm}

We are mainly interested in evaluating the reflected field in
medium 0, in a physical situation in which medium 3 is equal to
the medium 1, that is for $n_1=n_3$. This situation is related to
the possibility of using Bessel beams in searching for objects
buried underground as well as any kind of macro impurity or defect
inside a given material.

Because of the interference of multiple reflections on the layers,
the amplitude of the reflected field is an oscillating function
with respect to the frequency. For this reason, the intensity of
the beam greatly depends on the frequency. Thus, for given
parameter values, the choice of the frequency is extremely
important in order to have the Bessel beam amplitude surmounting
the amplitude of the plane wave.

In Fig. \ref{ro}  we show the intensity of the reflected field
(Eq. \ref{e0r}) from a Bessel beam and a plane wave as a function
of the radial coordinate $\rho$, for two different frequencies.
The figure is limited to the central portion of the Bessel beam,
since it is the only one of physical interest.

  In order to check the validity of the scalar approximation, for
the value of the parameter $\theta_0$ used here ($\theta_0
=30^\circ$), and for $\rho$ in the range of physical interest, we
have verified that one component of the electric field is much
larger than the other. The components of the field became
comparable only close to the first zero of the Bessel function,
and, in this case, the scalar approximation loses its validity.

The fields which describe the propagation of the plane wave
through the layers are given by Eqs. (\ref{e0i})-(\ref{e3}) by
putting $\theta_0 =0$.

 Our parameter values refer to the case of a sheet of glass buried
underground \cite{hip}. Near $\rho=0$ the gain is evident for both
frequencies. Even if the intensity of the reflected field is only
a few percent with respect to the incoming field \cite{note01},
the intensity of the Bessel beam in both cases is about three
times greater than that of the plane wave.

For complex refractive indices, we chose a mean standard value in
the microwaves range. Moreover, we set the value of the imaginary
part of the dielectric constant at approximately one order of
magnitude less than the real one. We are aware that this position
may appears to be rather rough. However, it is a plausible
approximation for our purposes, since our aim is to analyze the
difference in the propagation between a Bessel beam and a plane
wave, rather than to analyze a specific material (under specific
conditions of humidity, temperature etc.). We should recall that
the the complex refractive index $n_c=n_r+in_i$ is related to the
complex dielectric constant
$\epsilon_c=\epsilon^{'}+i\epsilon^{''}$ by means of the relation
$n_c^2=\epsilon_c$. Thus, for $\epsilon^{''}=0.1 \, \epsilon^{'}$
we simply have $n_i\simeq 0.05 \,\, n_r$.

Once the use of Bessel beams has been shown to be advantageous
with respect to that of plane waves, we can investigate the
difference of intensity in the reflected field in the presence or
absence of a given buried material.

\begin{figure}[t][h]
\includegraphics[scale=0.5]{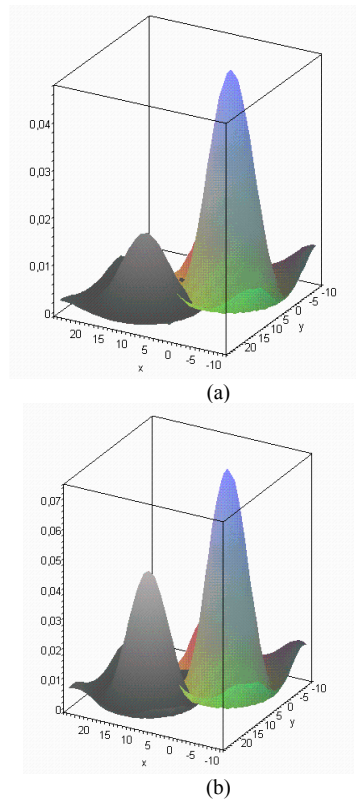}
\caption{Three-dimensional intensity of the reflected field (Eq.
\ref{e0r}) as a function of the Cartesian coordinates $x$ and $y$.
Higher signal refers to the reflected  field in the presence of
buried object; smaller signal (dark gray) refers to the
propagation through the only ground. Parameter values are:
$\theta_0=30^{\circ} , \: d_1=25$ cm, $d_2=1$ cm, and (a):
$n_{1r}=1.2,\: n_{2r}=2.25, \:\nu =2.2$ GHz; (b): $ n_{1r}=1.4,\:
n_{2r}=2.3, \:\nu =2.32$. We now need to make a comment or two on
the size of the Bessel beam generator: for $\theta_0= 30^\circ$,
in order to have a field depth of 50 cm, the converging system
must have have a radius of about 30 cm \cite{mug00}. This
dimension is not unusual for an electromagnetic mirror, while a
lens of this size would have to be custom-made. Naturally, the
size of the microwave generator-launcher is different, depending
on the frequency.} \label{map}
\end{figure}

In Fig. \ref{map} we report the three-dimensional intensity of the
reflected field as a function of the Cartesian coordinates $x$ and
$y$ $(\rho=\sqrt{x^2+y^2})$. The higher signal refers to the
passage in the presence of glass, while the lower one refers to
the passage only through the ground. The difference in the
intensities is evident so that, in an experimental investigation,
the presence of a buried object (glass, stone or ceramics
materials) should be clearly detected.

\vspace{1 cm}

We have shown that the use of Bessel beams can be advantageous as
compared to plane waves, in the detection of buried objects. Our
analysis refers to a particularly favorable case, that is, objects
buried inside very dry material (e.g. sandy or clayey ground)
\cite{sola}. In this case, in fact, microwaves undergo small
absorption.

 We showed that, near $\rho =0$, Bessel
beams always present a gain with respect to plane waves, provided
that a suitable frequency is chosen: the gain becomes more and
more negligible as the frequency increases. For high frequency
values, the advantage of using a Bessel beam is lost, since the
amplitude of the plane wave always exceeds that of the beam.

A second, and perhaps more important, aspect related to Bessel
beams is that the difference between the propagation through the
ground alone and the propagation through the ground in the
presence of buried objects is quite evident (Fig. \ref{map})
\cite{note2}.

In the light of these conclusions, we can think that a GPR (ground
penetrating radar) system which utilizes a Bessel beam as incoming
pulse could present some advantages as compared to a traditional
GPR apparatus.

Our theoretical model works under two main approximations: namely,
the smooth surfaces of the layers and homogeneous media. From an
experimental point of view, the first approximation works well as
long as the dimensions of roughness are sufficiently small as
compared with the wavelength; the second one requires a specific
software, in order to select between signal and noise.

Because of their feature of localized waves (i.e. localized
energy), Bessel beams could also provide further information about
the dimension and shape of buried objects. However, an analysis
devoted to this aspect goes beyond the purpose of the present work
and will be reported elsewhere.


\begin{thebibliography}{99}
\bibitem{note01}
The amplitude of the incoming field of the plane wave is constant
and equal to the unity. For the Bessel beam the amplitude is equal
to the unity only at $\rho =0$, and less to the unity elsewhere.
\bibitem{note2} For different materials, or different physical situations, this
difference changes and can also become very small.
\end{thebibliography}
\end{document}